\documentclass[preprintnumbers, floatfix,letterpaper,aps,prd,epsfig,nofootinbib,
twocolumn
]{revtex4-1}

\usepackage{bm,graphicx,dcolumn,epstopdf,epsf, latexsym,mathbbol, amssymb,amsmath,color,slashed, mathrsfs,mathcomp,simplewick}
\usepackage[center]{subfigure}
\usepackage{multirow}
\usepackage{makecell}
\usepackage[colorlinks,linkcolor=blue,citecolor=blue,urlcolor=blue]{hyperref}

\begin{document}
\allowdisplaybreaks
 \newcommand{\bq}{\begin{equation}}
 \newcommand{\eq}{\end{equation}}
 \newcommand{\bqn}{\begin{eqnarray}}
 \newcommand{\eqn}{\end{eqnarray}}
 \newcommand{\nb}{\nonumber}
 \newcommand{\lb}{\label}
 \newcommand{\f}{\frac}
 \newcommand{\p}{\partial}
\newcommand{\PRL}{Phys. Rev. Lett.}
\newcommand{\PLB}{Phys. Lett. B}
\newcommand{\PRD}{Phys. Rev. D}
\newcommand{\CQG}{Class. Quantum Grav.}
\newcommand{\JCAP}{J. Cosmol. Astropart. Phys.}
\newcommand{\JHEP}{J. High. Energy. Phys.}
\newcommand{\red}{\textcolor{red}}
%

\title{Black hole scalarizations induced by parity violations}

\author{Hao-Jie Lin${}^{a, b}$}
\email{haojie$\_$lin@zjut.edu.cn}

\author{Tao Zhu${}^{a, b}$}
\email{corresponding author: zhut05@zjut.edu.cn}

\author{Shao-Jun Zhang${}^{a, b}$}
\email{sjzhang@zjut.edu.cn}
\author{Anzhong Wang${}^{c}$}
\email{anzhong$\_$wang@baylor.edu}

\affiliation{
${}^{a}$ Institute for Theoretical Physics and Cosmology, Zhejiang University of Technology, Hangzhou, 310032, China\\
${}^{b}$ United Center for Gravitational Wave Physics (UCGWP), Zhejiang University of Technology, Hangzhou, 310032, China\\
${}^{c}$ GCAP-CASPER, Physics Department, Baylor University, Waco, Texas 76798-7316, USA}

\date{\today}

\begin{abstract}

It is well-known that parity symmetry is broken in the weak interaction but conserved for Einstein's general relativity and Maxwell's electromagnetic theory. Nevertheless, parity symmetry could also be violated in the gravitational/electromagnetic sectors if a fundamental scalar field couples to the parity-violating gravitational/electromagnetic curvature terms. Such parity-violating terms, which flip signs under reversed spatial directions, can inevitably lead to a negative effective mass squared for the scalar field perturbations near nonspherically symmetric black holes and thus are expected to trigger tachyonic instability. As illustrative examples, we show that the scalar field coupled to gravitational/electromagnetic Chern-Simons terms near a Kerr-Newmann spacetime can develop tachyonic instabilities, leading to equilibrium scalar field configurations in certain parameter regions of black holes. This instability, which is an indication of the black hole scalarization process, can occur in a broad class of nonspherically symmetric black holes and parity-violating theories. 

\end{abstract}

\maketitle

\section{Introduction}

In recent years, the availability of detection data on black holes has increased significantly \cite{a1, LIGOScientific:2021djp, a2, a3, a4}, making the study of black holes a topic of growing interest in the scientific community. It is widely accepted that, as an indispensable prediction of the theory of general relativity (GR), black holes can only be described in terms of their mass, electric charge, and angular momentum. Any other charge is not expected to exist according to the {\em no-hair theorem} \cite{Chrusciel:2012jk, Bekenstein:1998aw, Bambi:2017khi}. Remarkably, most of the current observations, including the detection of gravitational waves \cite{Isi:2019aib}, black hole images \cite{a2,a4}, and the stars orbiting the supermassive black hole in our Galactic Center \cite{Qi:2020xzi}, are in good agreement with hairless black holes. Future precise observations, such as gravitational wave detection and black hole images, can provide more accurate and details information about the nature of the black hole spacetime in the regime of strong gravity. Importantly, these precise observations can also provide a significant way to probe or constrain possible extra {\em hairs} in black holes. 

One type of such extra hairs, a fundamental scalar field, could survive around black hole spacetime in many classes of extended scalar-tensor theories through a tachyonic instability known as spontaneous scalarization. 
Hence, a growth of scalar hair or spontaneous scalarization can render the new fundamental scalar field detectable near black holes \cite{Maselli:2021men}.
Several mechanisms that account for the spontaneous scalarization around black holes have been proposed. In particular, matter-induced spontaneous scalarization was proposed for compact neutron stars in scalar-tensor theories \cite{ns1,ns2}. With nontrivial couplings of the scalar field to the spacetime curvature or electromagnetic field, the curvature-induced and charge-induced spontaneous scalarizations are also proven to be possible in the Einstein-Gauss-Bonnet gravity \cite{esgb1,esgb2,esgb3,esgb4,esgb5,Brihaye:2019kvj} and Einstein-scalar-Maxwell theories \cite{esm1,esm2,esm3,esm4}, respectively. More recently, it has been shown that the spontaneous scalarization induced by the spin of a black hole can also occur in various modified theories \cite{spin1,spin2,spin3,spin4,spin5,Annulli:2022ivr,Lai:2022spn}. In addition, the scalarization of black holes realized through the nonlinear accretion of scalar fields has also been discovered and has attracted a lot of attention recently \cite{nas1,nas2,nas3,nas4,nas5}. 

A key ingredient to trigger the spontaneous scalarization around a black hole is a threshold in both the coupling function, describing the interaction between the scalar field and gravitational/electromagnetic fields, and the black hole parameters, beyond which a tachyonic instability is induced by a sufficiently large and negative effective mass squared of the scalar field \cite{Doneva:2022ewd}. Such a negative effective mass squared sensitively depends on the sign of the coupling functions in both the curvature/charge-induced scalarization \cite{esgb3,esm1,esm2,esm3} and the value of the spin of the black hole in the spin-induced scalarization \cite{esgb5,esm3,esm4,spin1,spin2,spin3, Lai:2022spn}, for example. 

In this paper, we present a study on a mechanism for the spontaneous scalarization of nonspherically symmetric black holes, {\em the parity-violation-induced black hole scalarization}. It is worth mentioning that the spontaneous scalarization of black holes in a specific parity-violating theory---i.e., the Chern-Simons modified gravity---has been extensively studied previously \cite{p1,p2, Gao:2018acg,nl2, mdcsg, ecsg,rnmdcsg}. In these theories, the scalar field coupled with the parity-violating gravitational/electromagnetic field can inevitably lead to a negative effective mass squared for the scalar field and thus can trigger tachyonic instability when it is sufficiently large.

To illustrate this mechanism, we analyze the behavior of the scalar field which couples to two specific parity-violating terms: the gravitational and electromagnetic Chern-Simons terms, in a fixed Kerr-Newmann background. Due to the inherent complexity and nonlinearity of scalarization dynamics in rotating black hole backgrounds, our study only focuses on the scalarization dynamics within the framework of the ``decoupling limit." In this way, we numerically evolve the nonlinear scalar field equation on the fixed Kerr-Newmann geometry, disregarding the backreaction of the scalar field on the background spacetime \cite{nl2}. This simplification allows us to gain valuable insights into the scalarization phenomenon while bypassing the significant challenges associated with capturing its complete generality and nonlinearity. Our result reveals that the scalar field in a Kerr-Newmann spacetime can develop tachyonic instabilities, eventually leading to the formation of an equilibrium scalar field configuration. 

We additionally check that the effective mass squared of the scalar field can be inevitably negative in a broad class of nonspherically symmetric black holes and parity-violating theories, which are expected to lead to tachyonic instabilities in specific regions of the spacetime.  Our results suggest that the phenomenon of black hole scalarization potentially occurs for a broad class of nonspherically symmetric black holes in numerous parity-violating theories.

\section{The model}

Let us describe our model by starting with the following scalar field action on a nontrivial black hole spacetime, 
\begin{eqnarray}
S_\phi = - \int d^4 x \sqrt{-g}\left(\frac{1}{2}\nabla_\mu \phi \nabla^\mu \phi + f(\phi) I(\psi; g_{\mu\nu})\right),\nb\\
\end{eqnarray}
where $\phi$ is the scalar field, $I(\psi; g_{\mu\nu})$ represents the source term depending on $g_{\mu\nu}$ and matter fields $\psi$, and $f(\phi)$ is the coupling function determining the coupling strength between the scalar field $\phi$ and the spacetime metric $g_{\mu\nu}$ and other matter fields $\psi$. By varying the above action with respect to the scalar field one obtains the equation of motion of the scalar field on the black hole spacetime,
\begin{eqnarray}
\Box \phi - f'(\phi) I=0,\lb{seq0}
\end{eqnarray}
where $f'(\phi)=df(\phi)/d\phi$. This equation allows for the existence of scalar-free solutions which are also solutions of GR with conditions $\phi=0$ and $f'(0)=0$ \cite{Astefanesei:2020qxk,Herdeiro:2021ftk}. And spontaneous scalarization occurs if the scalar-free solution is unstable against scalar perturbations $\delta \phi$. To study how the scalar-free background is affected by the small scalar field perturbation $\delta\phi$, it is convenient to consider the linearized scalar field equation, which is
\bqn
(\Box - \mu_{\rm eff}^2)\delta \phi=0,\lb{seq}
\eqn
where $\mu_{\rm eff}^2=f''(0) I(\psi; g_{\mu\nu})$ is the effective mass squared of the scalar field. Once $\mu^2_{\rm eff}$ becomes sufficiently negative, tachyonic instability occurs, causing the scalar-free spacetime to be unstable under $\delta \phi$ in a certain region of the parameter space, indicating an onset of the black hole scalarization process \cite{Doneva:2022ewd}. 

One well-studied model of the spontaneous scalarization is realized in the framework of the Einstein-scalar-Gauss-Bonent gravity with $I(\psi; g_{\mu\nu})=R_{\rm GB}^2$, where $R_{\rm GB}^2$ denotes the Gauss-Bonent scalar. For a Schwarzschild black hole, one has $R_{\rm GB}^2 = 48 M^2/r^6$, with $M$ being the mass of the black hole. It is evident that effective mass $\mu_{\rm eff}$ is only allowed to be negative for $f''(0)<0$, with which a tachyonic instability occurs when $|f''(0)|$ is sufficiently large \cite{spin3}. Similar curvature-induced scalarization occurs in Reissner-Nordström(RN) and Kerr black holes as well with a negative $f''(0)$ \cite{esgb2,Brihaye:2019kvj,spin3,spin4,spin5}. For a Kerr black hole, $\mu_{\rm eff}^2$ can become negative even for positive $f''(0)$ with a high spin and thus can trigger the spin-induced scalarization \cite{spin3,spin4,spin5}. The black hole scalarization with other models---for example, $I(\psi, g_{\mu\nu})=F_{\mu\nu}F^{\mu\nu}$---have also been extensively studied in the literature \cite{esm1,esm2,esm3,esm4,Lai:2022spn}. 

In this paper, we consider two choices of $I(\psi, g_{\mu\nu})$:
\bqn
I(\psi, g_{\mu\nu}) = R \tilde R\;\;\;{\rm and} \;\;\; I(\psi, g_{\mu\nu}) =F \tilde F,
\eqn
representing two parity-violating interactions between the scalar field and the gravitational/electromagnetic field, where $R\tilde R=\frac{1}{2}\epsilon^{\mu\nu\lambda\gamma} R^{\eta}_{~\xi \lambda \gamma}R^{\xi}_{~\eta \mu \nu}$ and $F\tilde{F}=\frac{1}{2}\epsilon^{\mu \nu \lambda \gamma}F_{\mu \nu}F_{\lambda \gamma}$ with $\epsilon^{\mu\nu\lambda\gamma}$ being the totally antisymmetric Levi-Civita tensor. Intriguingly, such couplings break the parity symmetry in the gravitational/electromagnetic sectors, and thus $I(\psi; g_{\mu\nu}) \to - I(\psi; g_{\mu\nu})$ under parity transformations. When $ I(\psi; g_{\mu\nu})$ does not vanish for a nontrivial black hole background, $I(\psi; g_{\mu\nu})$ has to be negative in a certain region of the spacetime, which indicates that a negative effective mass squared for the scalar field inevitably exists. When $\mu_{\rm eff}^2 = f''(0) I(\psi; g_{\mu\nu})$ is sufficiently large and negative, exceeding a threshold, the tachyonic instability of the scalar field occurs which triggers the process of the spontaneous scalarization.

\section{Scalarization process in Kerr-Newmann black holes}

To illustrate the mechanisms of the parity-violation-induced scalarization process, let us consider the scalarization of the Kerr-Newmann black hole as a concrete example, in which both of the parity-violating terms $R\tilde R$ and $F\tilde{F}$ do not vanish. 

In the Boyer-Lindquist coordinates $\{t,r,\theta,\varphi\}$, the Kerr-Newmann metric with charge $Q$, mass $M$, and spin $a$, is given by 
\bqn
d s^2 &\equiv&-\frac{\Delta-a^2 \sin ^2 \theta}{\Sigma^2} d t^2-\frac{2 a \sin ^2 \theta\left(r^2+a^2-\Delta\right)}{\Sigma^2} d t d \varphi \nb\\
&&+\frac{\left[\left(r^2+a^2\right)^2-\Delta a^2 \sin ^2 \theta\right] \sin ^2 \theta}{\Sigma^2} d \varphi^2\nb\\
&&+\frac{\Sigma^2}{\Delta} d r^2+\Sigma^2 d \theta^2,
\eqn
where
\bqn
\Delta&=&r^2-2 M r+a^2+Q^2, \\ \Sigma^2&=&r^2+a^2 \cos ^2 \theta,\\a&=&\frac{J}{M} .
\eqn
The corresponding vector potential is
\bqn
A_{\mu}dx^{\mu}=-\frac{Q r}{\Sigma^2}\left(d t-a \sin ^2 \theta d \varphi\right).
\eqn
Therefore, the explicit form of the parity-violating terms $R\tilde R$ and $F\tilde{F}$ in a Kerr-Newmann black hole, are given, respectively, by 
\bqn
R \tilde R &=& \frac{96 a \cos \theta\left[\left(3 M r-2 Q^2\right) r-M a^2 \cos ^2 \theta\right]}{\left(r^2+a^2 \cos ^2 \theta\right)^6}\nb\\
&& \times \left[\left(M r-Q^2\right) r^2-\left(3 M r-Q^2\right) a^2 \cos ^2 \theta\right], \nb\\
\\
F\tilde{F}&=&\frac{8 a Q^2 r \cos \theta\left(r^2-a^2 \cos \theta^2\right)}{\left(r^2+a^2 \cos \theta^2\right)^4}.
\eqn
Intriguingly, both of the parity-violating terms contain a factor $\cos\theta$, such that it always has a negative sign under the following parity transformation: 
\bqn
&&\theta \rightarrow \pi -\theta, \\
&&R \tilde R \to - R \tilde R, \\
&&F\tilde{F}\rightarrow -F\tilde{F},
\eqn
which implies that regardless of the specific form of $f(\phi)$, the effective mass square $\mu_{\rm eff}^{2}$ always has negative values in the interval $\theta \in [0, \pi]$. Therefore, this nonminimal coupling between the parity-violating terms and the scalar field will inevitably lead to tachyonic instability as long as $|f''(0)|$ is sufficiently large, resulting in the occurrence of the spontaneous scalarization process of Kerr-Newman black holes at the linear level.

To further validate this assertion, we consider the specific choice, $f(\phi) = \frac{\alpha}{2\beta}(1-e^{-\beta \phi^2})$ \cite{nl1,nl2}, where $\alpha$ and $\beta$ are two constants. When $\beta\rightarrow 0$, the coupling function $f(\phi)$ reduces to the quadratic form $\frac{1}{2}\alpha \phi^2$, which is sufficient for studying spontaneous scalarization resulting from linear tachyonic instability. As the instability progresses and the scalar field grows, the significance of nonlinear terms increases and eventually quenches the instability. The final state is an equilibrium scalar field configuration.

To solve Eq.~(\ref{seq0}) for the time evolution of the scalar field, we utilize a hyperboloidal foliation method described in Ref. \cite{Zhang:2020pko}. 

First, we transform the Bayer-Lindquist coordinates into the ingoing Kerr-Schild coordinates $\{\tilde{t},r,\theta ,\tilde{\varphi} \}$:
\bqn
	d \tilde{t}&=&d t+\frac{2 M r-Q^2}{\Delta} d r, \\
	d \tilde{\varphi}&=&d \varphi+\frac{a}{\Delta} d r .
\eqn
Considering the axisymmetry of the Kerr-Newmann spacetime, the scalar perturbation can be decomposed as 
\bqn
\phi(\tilde{t}, r, \theta, \tilde{\varphi})=\sum_m \frac{\Psi_m(\tilde{t}, r, \theta)}{r} e^{i m \tilde{\varphi}},
\eqn
where $m$ is the azimuthal mode number and $\Psi_m$ is a new field variable.\footnote{The introduction of a conformal compactification, as described in Eqs. (\ref{s1}) and (\ref{s2}), necessitates the rescaling of the scalar field variable, such as $\phi=r^{-1}\Psi$. This rescaling is crucial to mitigate the singularity of the physical metric at the future null infinity $\mathscr{I}^{+}$, which is also present in the wave equation \cite{te1,te2}.}
Substituting the above expressions into Eq.~(\ref{seq0}), we obtain
\bqn\lb{pde1}
&&A^{\tilde{t} \tilde{t}} \partial_{\tilde{t}}^{2} \Psi_m+A^{\tilde{t} r} \partial_{\tilde{t}} \partial_{r} \Psi_m+A^{r r} \partial_{r}^{2} \Psi_m+A^{\theta \theta} \partial_{\theta}^{2} \Psi_m \nb\\
&&+B^{\tilde{t}} \partial_{\tilde{t}} \Psi_m+B^{r} \partial_{r} \Psi_m+B^{\theta} \partial_{\theta} \Psi_m+C \Psi_m=0,\nb\\
\eqn
where
\bqn
&&A^{\tilde{t} \tilde{t}} =2 M r-Q^2+\Sigma^{2}, \nb\\
&&A^{\tilde{t r}} =-4 M r+2Q^2, \nb\\
&&A^{r r} =-\Delta, \nb\\
&&A^{\theta \theta} =-1 ,\nb\\
&&B^{\tilde{t}}= 2 M-\frac{2Q^2}{r} ,\nb\\
&&B^{r} =\frac{2\left(a^{2}+Q^2-M r\right)}{r}-2 i m a, \\
 &&B^{\theta}=-\cot \theta, \nb \\
&&C=-\frac{2\left(a^{2}+Q^2-M r\right)}{r^{2}}+\frac{2 i m a}{r}+\frac{m^{2}}{\sin ^{2} \theta}\nb\\
 &&\quad\quad+\Sigma^{2} \alpha e^{-\beta(\sum_m\Psi_m e^{i m \tilde{\varphi}}/r)^2}I.\nb
\eqn
Each azimuthal mode $m\neq 0$ has harmonic dependence on $\tilde{\varphi}$, which causes Eq.~(\ref{pde1}) to be nonlinear. To eliminate the interference from $\tilde{\varphi}$, we only consider the axisymmetric mode with $m = 0$ to study the time evolution of the scalar field perturbation. This particular choice has an impact on the excited quasinormal modes in stable scenarios and the characteristic times involved, but it does not alter the overall outcome of the scalarization process \cite{Doneva:2022yqu}.

The second step is to define the compactified horizon-penetrating, hyperboloidal coordinates (HH coordinates) $\{\tau, \rho, \theta, \tilde{\varphi}\}$. Specifically, we replace the ingoing Kerr-Schild coordinates $\tilde{t}$ and $r$ with 
\bqn
\tilde{t}&=&\tau+h(\rho), \lb{s1}\\
 r&=&\frac{\rho}{\Omega(\rho)},\lb{s2}
\eqn
where 
\bqn
h(\rho)&=&\frac{\rho}{\Omega}-\rho-4M \mathrm{ln}\Omega,\\
\Omega(\rho)&=&1-\frac{\rho}{S}.
\eqn

By these coordinate transformations, the future null infinity $\mathscr{I}^{+}$ is compactified at $\rho=S$, and the event horizon of the Kerr-Newman black hole $r_+=M+\sqrt{M^2-a^2-Q^2}$ is located at
\bqn
\rho_+=\frac{S^2\left(M+\sqrt{M^2- a^2-Q^2}\right)+S \left(a^2+Q^2\right) }{a^2+2 M S+Q^2+S^2}.\nb\\
\eqn
Next, we define a boost function as 
\bqn
H(\rho)=\frac{dh(\rho)}{dr},
\eqn
and then the partial derivatives for $\tilde{t}$ and $r$ can be rewritten as
\bqn
 \partial_{\tilde{t}}=\partial_{\tau},~\partial_{r}=-H \partial_{\tau}+\frac{d \rho}{d r} \partial_{\rho}.
\eqn
Substituting the above expressions into Eq.~(\ref{pde1}), we obtain
\bqn\label{pde2}
 &&A^{\tau\tau} \partial_{\tau}^{2} \Psi_m+A^{\tau \rho} \partial_{\tau} \partial_{\rho} \Psi_m+A^{\rho\rho} \partial_{\rho}^{2} \Psi_m+A^{\theta \theta} \partial_{\theta}^{2} \Psi_m\nb\\
 &&+B^{\tau} \partial_{\tau} \Psi_m+ B^{\rho} \partial_{\rho} \Psi_m+B^{\theta} \partial_{\theta} \Psi_m+C\Psi_m=0~,\nb\\
\eqn
where
\bqn
 &&A^{\tau \tau}=A^{\tilde{t} t}-H A^{\tilde{t r}}+H^{2} A^{r r}, \nb\\
 &&A^{\rho \rho}=\left(\frac{d \rho}{d r}\right)^{2} A^{r r},\nb\\
 &&A^{\tau \rho}=\frac{d \rho}{d r}\left(A^{\tilde {t}r}-2 H A^{r r}\right), \\
 &&B^{\tau}=B^{\tilde{t}}-H B^{r}-\frac{d H}{d r} A^{r r} ,\nb\\
 &&B^{\rho}=\frac{d \rho}{d r}\left[B^{r}+\frac{d}{d\rho}\left(\frac{d\rho}{d r}\right) A^{r r}\right] ,\nb
\eqn
By introducing a new auxiliary variable $\Pi_{m} =\partial_{\tau}\Psi_m $, Eq.(\ref{pde2}) is recast into a first-order form of coupled partial differential equations:
\bqn
 \partial_{\tau}\Psi_m&=&\Pi_{m}, \\
 \partial_{\tau} \Pi_{m}&=&-\frac{1}{A^{\tau \tau}}\left(A^{\tau \rho} \partial_{\rho} \Pi_{m}+A^{\rho\rho} \partial_{\rho}^2 \Psi_m+A^{\theta \theta} \partial_{\theta}^{2}\Psi_m\right.\nb\\
 &&\left.+B^{\tau} \Pi_{m}+B^{\rho} \partial_{\rho}\Psi_m+B^{\theta} \partial_{\theta}\Psi_m+C\Psi_m\right).\nb\\
\eqn
 In terms of numerical implementations, we employ a fourth-order finite difference method for spatial grid discretizations, while the time evolution is accomplished using a fourth-order Runge-Kutta integrator. The HH coordinates automatically satisfy the ingoing/outgoing boundary condition at the horizon/infinity, eliminating the need to handle the complex outer boundary problem that can impact accuracy. However, at the angular poles ($\theta=0$ and $\pi$), we impose physical boundary conditions: $\left.\Psi_m\right|_{\theta=0, \pi}=0$ for $m\neq 0$, and $\left.\partial_{\theta} \Psi_m\right|_{\theta=0, \pi}=0$ for $m=0$ \cite{bd}. For the initial data, we consider a Gaussian distribution localized outside the horizon at $\rho = \rho_c$. Specifically, we have 
 \bqn
 &&\Psi_{lm}(\tau=0,\rho,\theta)\sim \mathbf{Y}_{lm}(\theta)e^{-\frac{\left(\rho-\rho_c\right)^2 }{2\sigma}} \\
 &&\Pi_m(\tau=0,\rho,\theta)=0
 \eqn
where $\mathbf{Y}_{lm}$ represents the $\theta$-dependent part of the spherical harmonic function, and $\sigma$ represents the width of the Gaussian distribution. 

It is noteworthy to mention that the absence of spherical symmetry in rotating black holes, leads to a phenomenon known as mode mixing. This phenomenon has been extensively studied in Kerr and Kerr-Newmann black holes in different theories of gravity, including GR \cite{mx1, mx2, mx3}, Chern-Simons modified gravity \cite{ecsg,Gao:2018acg}, Gauss-Bonnet gravity \cite{spin3,spin5,esgb5}, and Einstein-Maxwell-scalar gravity \cite{Lai:2022spn}, which indicates that a pure initial $l$ multipole will induce the presence of other $l'$ multipoles with the same $m$ as it evolves. The evolution of different $m$ modes is decoupled in the presence of axisymmetry. Additionally, the reflection symmetry results in the decoupling of even $ l$ modes from odd $l$ modes. However, the evolution of a specific mode $(l, m)$ is coupled to that of all the modes $(l + 2k, m)$ with $k$ being an integer \cite{spin3}. Notably, the $l = |m|$ mode plays a prominent role in the later stages of the evolution of the scalar field, 
as indicated in Ref.~\cite{spin5}. Among the dominant modes with different values of $m$, the mode with $m = 0$ exhibits the shortest growth times, holding the greatest relevance to the instabilities, and has the largest parameter region of the instability of the Kerr/Kerr-Newmann black holes \cite{spin5, Gao:2018acg, spin3, esgb5}. 
Moreover, the equations governing the evolution of the scalar field perturbation with the chosen azimuthal mode, $m = 0$, are linear in nature. This will significantly simplify the analysis as we avoid the complexities associated with solving nonlinear partial differential equations. Therefore, we solely focus on the perturbations with $l=m=0$ in the subsequent discussion. Additionally, we adopt the convention of setting $M = 1$ to express all quantities in units of $M$.

\begin{figure}
{
\includegraphics[width=8cm]{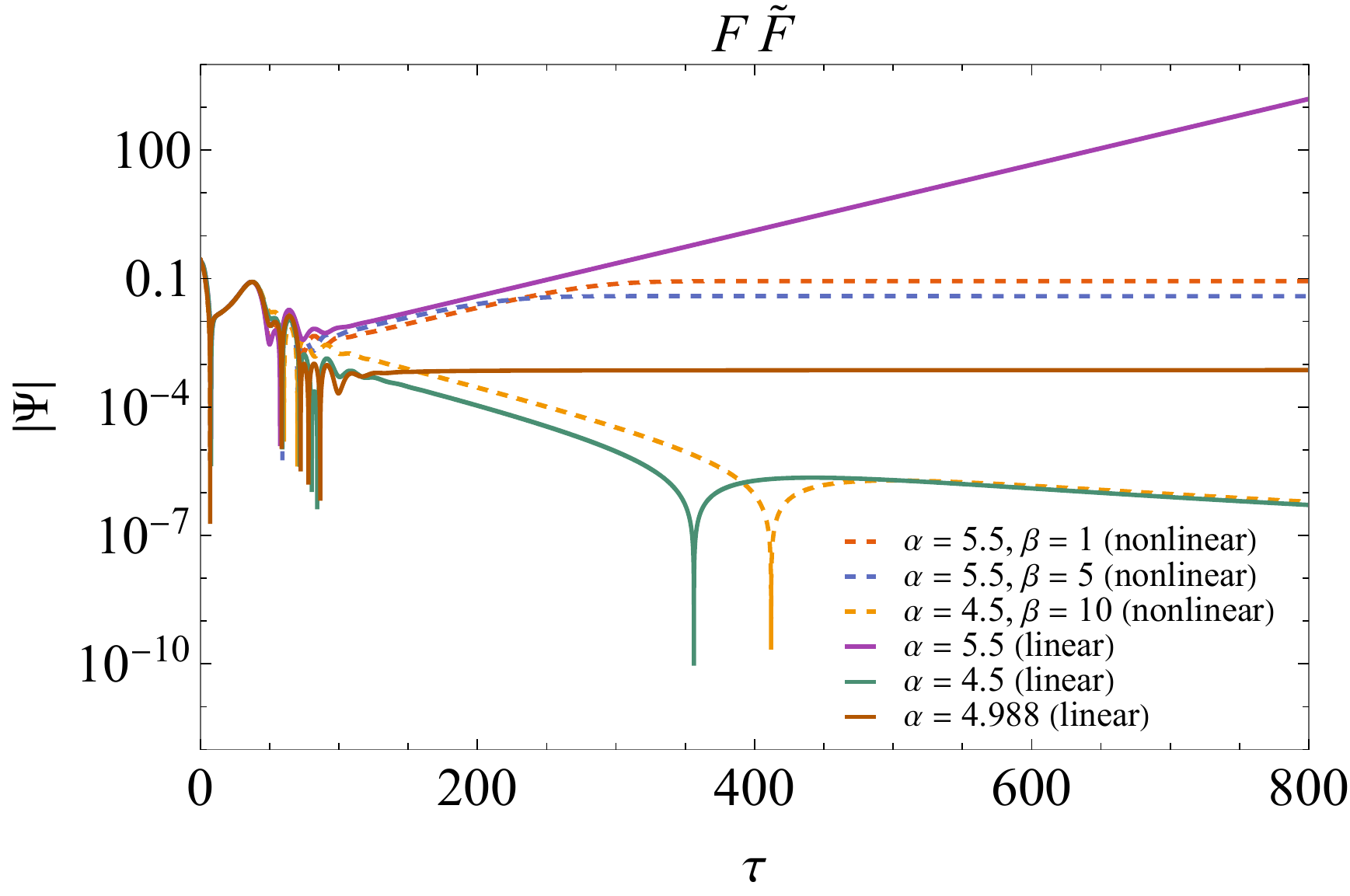}
\includegraphics[width=8cm]{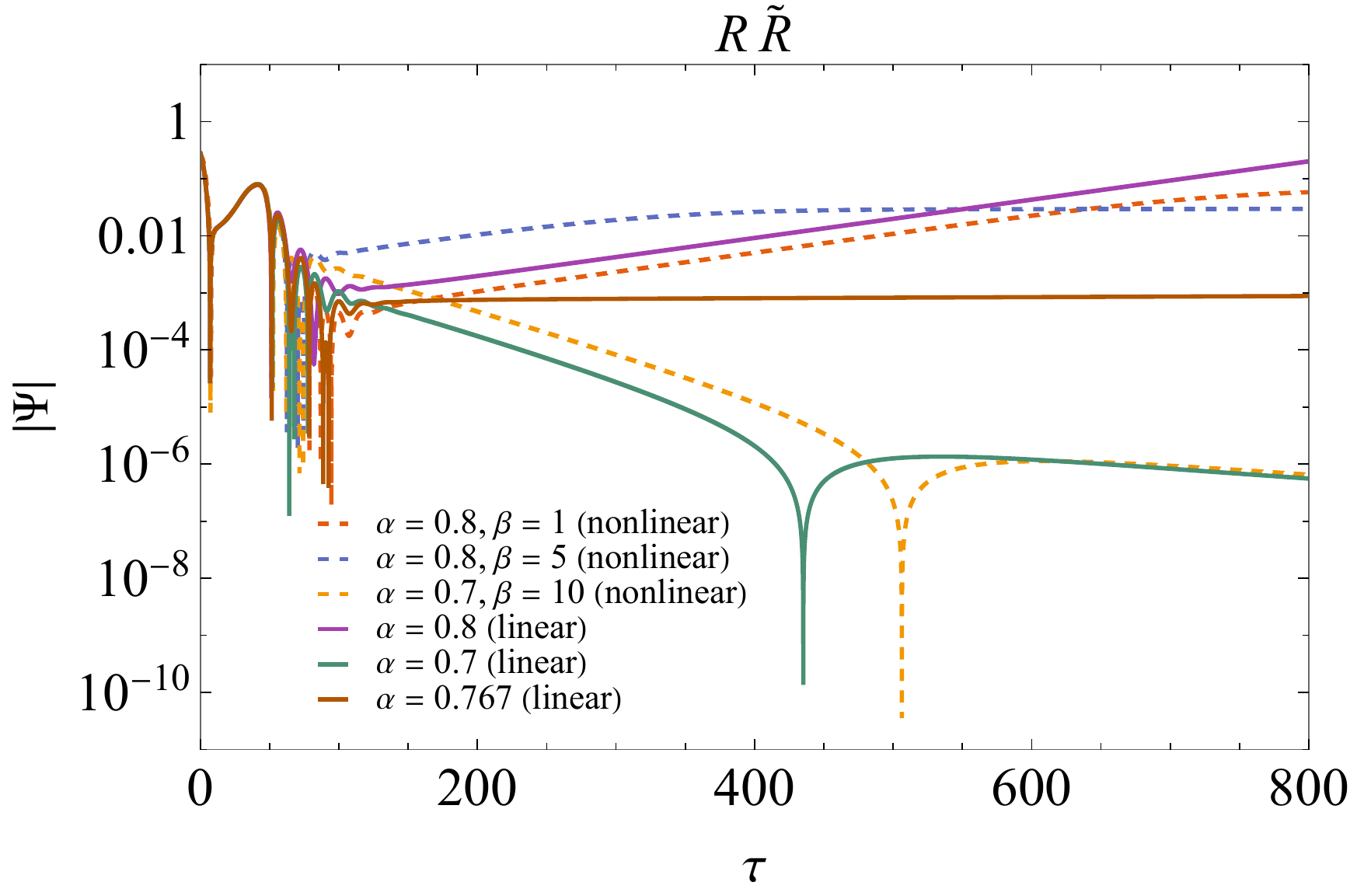}
}
\caption{Time evolutions of the scalar field perturbation due to $F\tilde{F}$ and $R\tilde{R}$ for $a=0.5$, $Q=0.8$ with various values of $\alpha$ and $\beta$.
Here we fix $\rho_c=6$ and $\sigma=0.2$.
Observers are assumed to be located at  $\rho=6$ and $\theta = \pi/4$.} 
\label{ax}
\end{figure}

\begin{figure}
{
\includegraphics[width=8cm]{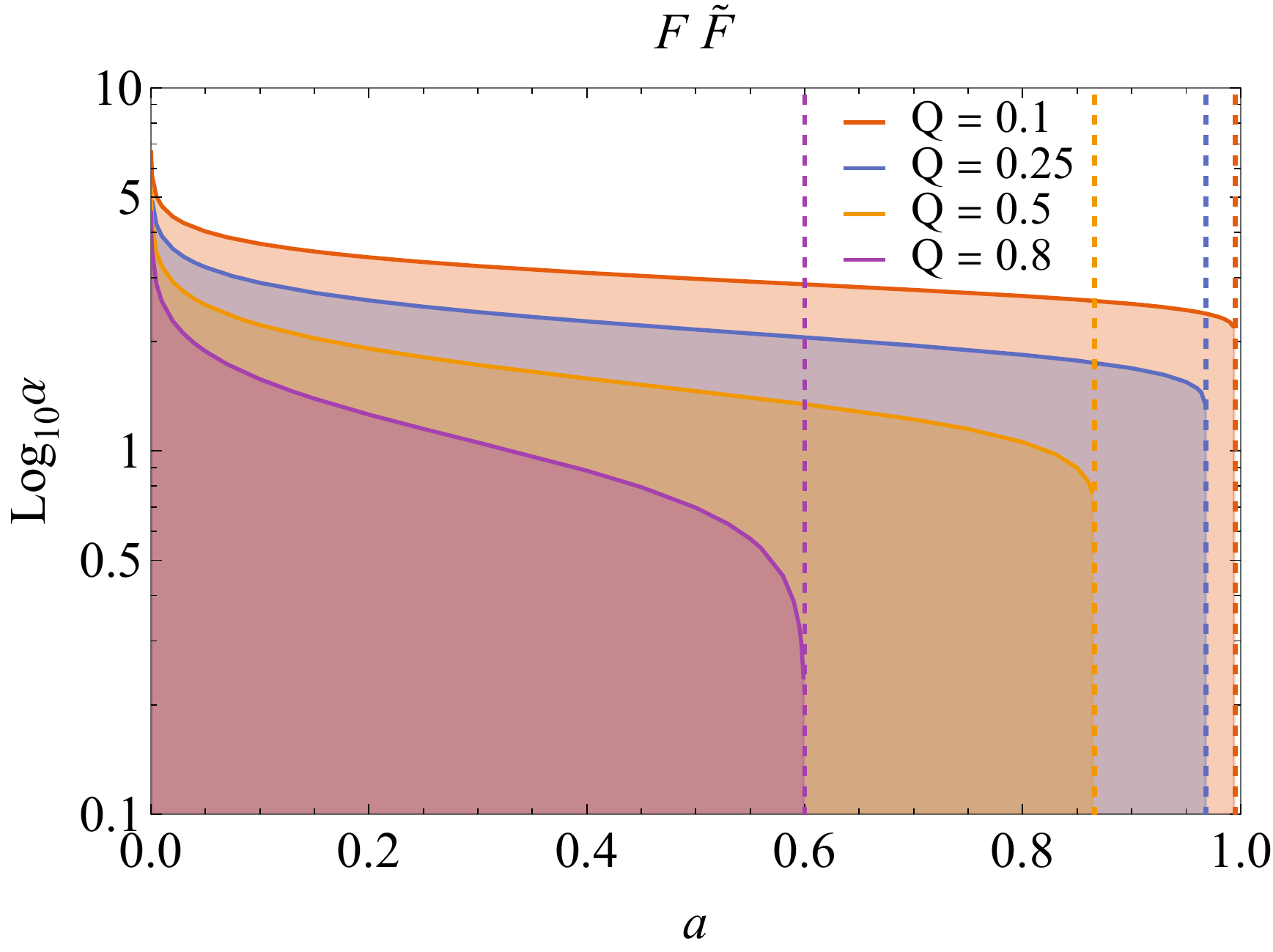}
\includegraphics[width=8cm]{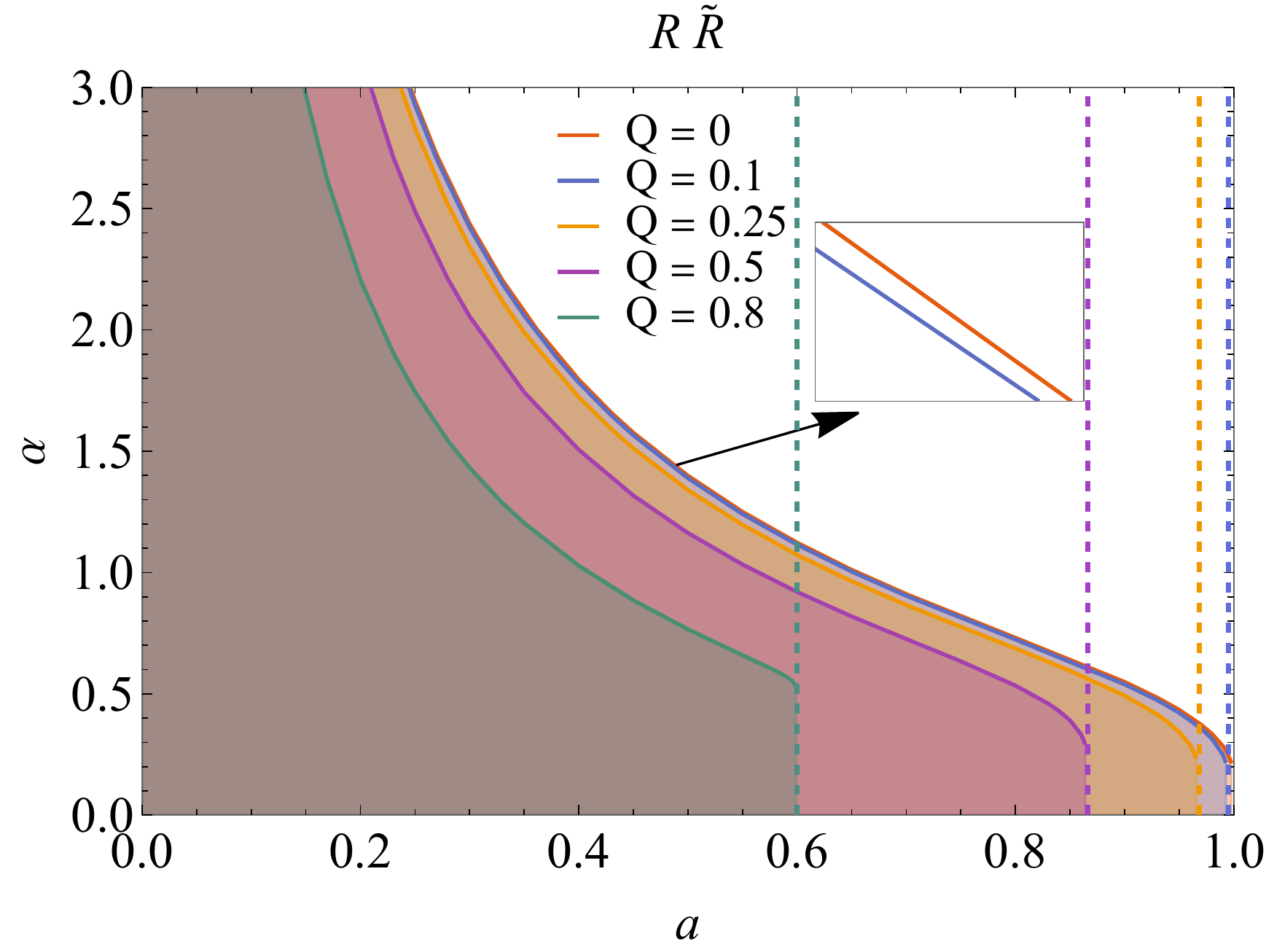}}
\caption{The parameter spaces of $a$ and $\alpha$ for $F\tilde{F}$ and $R\tilde{R}$ at different $Q$ values ($0.8$, $0.5$, $0.25$, and $0.1$) in the linear level. The solid lines represent the thresholds of the dominant $m=0$ mode for which scalarization occurs. The dashed lines represent the upper limits for the spin parameter for the fixed charges $Q$ (extreme Kerr-Newmann black holes). The filled areas indicate the stable regions, while the areas between the dashed and solid lines of the same color in both panels indicate instability. }\label{para}
\end{figure}

Our results, as presented below, demonstrate that the occurrence of instability at the linear level is contingent upon the values of spin $a$, charge $Q$, and $\alpha$. Figure \ref{ax} illustrates the time evolution of the scalar field for a fixed set of parameters. It is evident that the scalar field decays over time when $\alpha$ is below the scalarization threshold. However, when $\alpha$ surpasses this threshold, the scalar field experiences exponential growth. As anticipated, such growth is quenched by nonlinearity of the scalar field, leading to an equilibrium scalar field configuration, i.e., the black hole scalarizes.
Conversely, in the case of a  decaying scalar field perturbation, the effect of nonlinearity is largely insignificant.
Figure \ref{para} depicts the parameter space in which spontaneous scalarization arises due to the linear tachyonic instability, where the extreme Kerr-Newmann black hole provides an upper limit for the spin parameter for a fixed charge $Q$ (dashed lines). In the case of zero spin, the Kerr-Newmann black hole degenerates to a RN black hole with no parity-violating terms, leading to a decaying scalar field perturbation. For  $Q = 0$,  $F\tilde{F}$ vanishes, and only $R\tilde{R}$ remains, which has been extensively investigated in Refs.~\cite{Gao:2018acg,nl2}. Additionally, as $Q$ increases, the parameter space for the linear stability of the scalar field decreases.

\section{Extensions to other nonspherically symmetric black holes}

The scalarizations of the Kerr-Newmann black hole, with $I(\psi; g_{\mu\nu})$ being chosen to be the gravitational/electromagnetic Chern-Simons term, represent two illustrative examples of the scalarized mechanism induced by the parity violation. One remarkable property of such a mechanism is that it can exist in a broad class of black hole backgrounds, provided that the gravitational/electromagnetic Chern-Simons term does not vanish. Due to the parity violations, $I(\psi; g_{\mu\nu})$ simply vanishes in the spherical backgrounds and could be nontrivial only for nonspherically symmetric black holes.

\begin{figure}
\includegraphics[width=8cm]{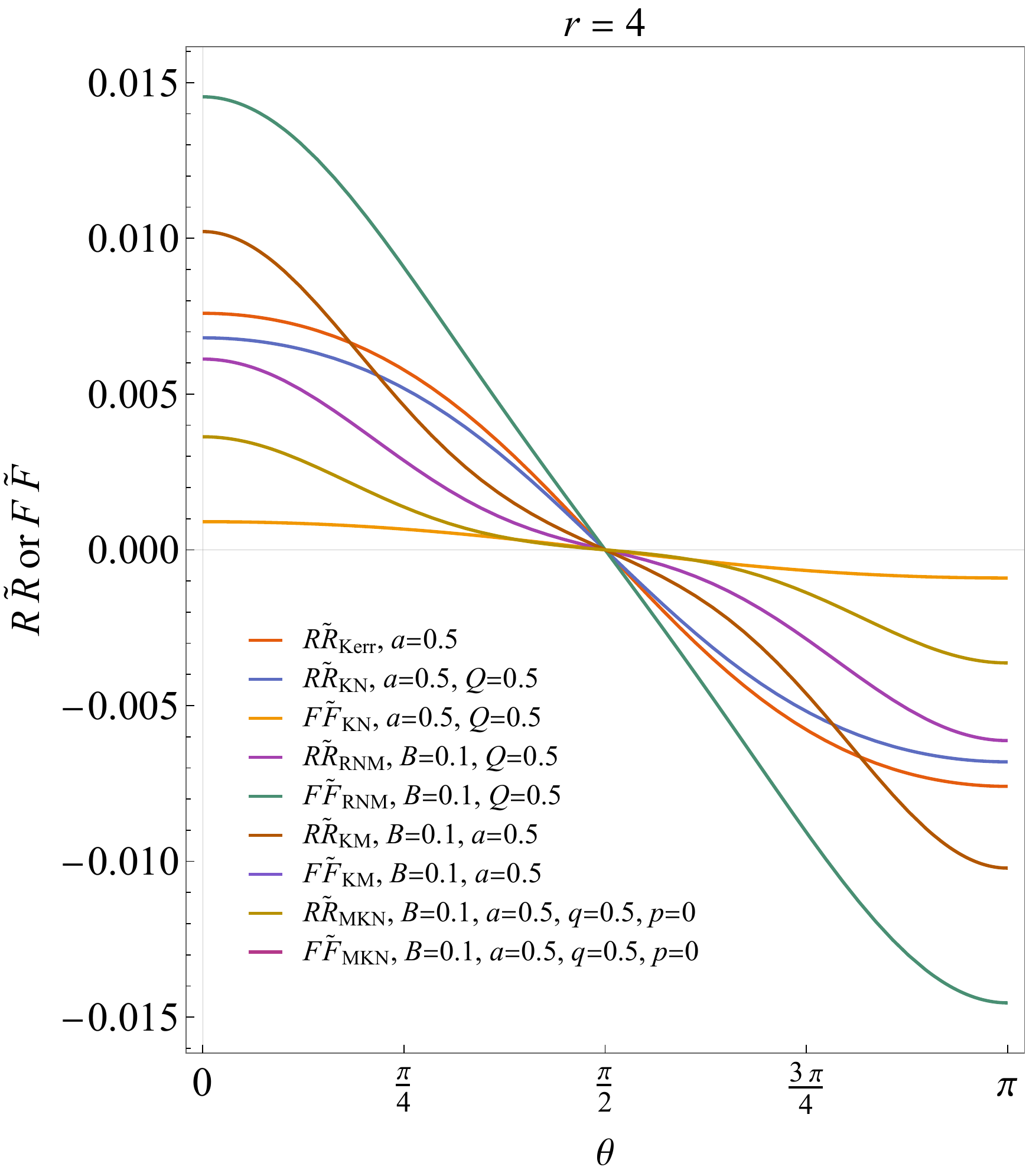}
\caption{The parity-violating terms of non-spherical black holes as a function of $\theta$, where $B$ represents the magnetic field strength, $Q$ represents the charges of the KN and RNM black holes, and $q$ and $p$ represent the electric and magnetic charges of the MKN black holes, respectively. We simply choose $r=4$. For other value of $r$, the result is similar.} \label{rnm}
\end{figure}

We now show the nonzero of the parity-violating gravitational/electromagnetic curvature invariants, $R\tilde R$ and $F  \tilde F$ for several specific nonspherically symmetric black holes. We consider five different nonspherically symmetric black hole solutions, including the Kerr, Kerr-Newmann, RN-Melvin~\cite{rnm1,rnm2}, Kerr-Melvin~\cite{km1}, and Magnetized Kerr-Newmann black holes~\cite{rnm1,mkn1,km1,rnm2}. As shown in Table~\ref{RRFF}, $R\tilde R$ is nonzero for all the five black holes, while $F\tilde F$ is nonzero for four spacetimes, all except the Kerr black hole. Once $R \tilde R$ or $F \tilde F$ is nonzero, it must give rise to a negative effective mass squared in a certain region of spacetime for the scalar perturbation no matter what the sign of $f''(0)$ is. This is shown in Figure~\ref{rnm}, in which we find the $R\tilde R$ and $F \tilde F$ for all the nonspherically symmetric black holes considered in this paper. It is evident that all the nonzero $R\tilde R$ and $F \tilde F$ exhibit a negative sign in the region $\theta \in [0, \pi]$ outside the horizon of the nonspherically symmetric black holes. When this negative mass squared is sufficiently negative, the scalar field will develop a tachyonic instability, resulting in the spontaneous scalarization of the nonspherically symmetric black holes. Here we would like to mention that the spontaneous scalarization of the Kerr and RN-Melvin black holes with $I=R \tilde R$ has been previously investigated in Refs.~\cite{p2, Gao:2018acg,nl2, mdcsg, ecsg,rnmdcsg}.

\begin{table}
\caption{\label{RRFF}%
Summary of the nonzero of $R \tilde R$ and $F \tilde F$ for several specific nonshperical black holes. Here a ``\checkmark" represents the nonzero of $R \tilde R$ or $F \tilde F$. }
\begin{ruledtabular}
\begin{tabular}{ccc}
  & \multicolumn{2}{c}{$I(\psi; g_{\mu\nu})$} \\
\cline{2-3} 
Black holes  & $R \tilde R$ & $F \tilde F$  \\
  \colrule
Kerr&  \checkmark & ... \\
Kerr-Newmann (KN) & \checkmark & \checkmark \\
Reissner-Nordström-Melvin (RNM) &  \checkmark & \checkmark \\
Kerr-Melvin (KM) &  \checkmark & \checkmark  \\
Magnetized Kerr-Newmann (MKN) &  \checkmark & \checkmark \\
\end{tabular}
\end{ruledtabular}
\end{table}

\section{Extensions to other parity-violating theories}

The models studied in this paper can be easily extended to other types of parity-violating theories---for example, the chiral-scalar tensor theory which extends the Chern-Simons gravity by including parity-violating interactions between the higher derivatives of the scalar field and the spacetime curvatures \cite{Crisostomi:2017ugk}. Similarly to the Chern-Simons gravity, the chiral-scalar-tensor theory can naturally trigger the tachyonic instability of the scalar field in non-spherical backgrounds and can exhibit more rich phenomena of black hole scalarizations. 

In the framework of the teleparallel gravity, the GR equivalent teleparallel gravity can be modified by adding a parity-violating interaction between the scalar field and the Nieh-Yan term in the gravitational action, i.e., $I(\psi; g_{\mu\nu})= {\cal T}_{A\mu\nu} \tilde {\cal T}^{A\mu\nu}$ with ${\cal T}_{A\mu\nu}$ being the torsion tensor two-form and $\tilde {\cal T}^{A\mu\nu} = \frac{1}{2}\epsilon^{\mu\nu\lambda \gamma} {\cal T}^A_{~~\lambda \gamma}$ being the dual torsion tensor two-form \cite{Li:2020xjt, Li:2021wij}. Similarly, the parity-violating symmetric teleparallel gravity can also be constructed by including an interaction between the scalar field and the nonmetricity tensor $Q_{\mu\nu \lambda}$---i.e., $I(\psi; g_{\mu\nu})=\epsilon^{\mu\nu\lambda\gamma} Q_{\mu\nu}^\xi Q_{\lambda \gamma \xi}$  \cite{Li:2021mdp}. More parity-violating gravities can be found in Refs.~\cite{Conroy:2019ibo, Rao:2023doc, Chen:2022wtz} and references therein. 

In all these theories, $f''(0)I(\psi; g_{\mu\nu})$ can inevitably become negative in a certain region of the nontrivial black hole backgrounds and thus can naturally trigger tachyonic instabilities. Therefore, it is interesting to explore such parity-violation-induced black hole scalarization and the corresponding scalarized black hole solutions in such a broad class of theories. We would like to address this issue in detail in our future works. 

\section{Summary and Discussion}

The results presented here exhibit a possible black hole scalarization process induced by the couplings between a scalar field and the parity-violating spacetime/electromagnetic curvature terms, potentially extending the family of known black hole scalarization mechanisms in the literature to a broader range of modified theories of gravity.

For illustrative purposes, we only consider the dynamics in the ``decoupling limit" by numerically evolving the nonlinear scalar field equation on the fixed Kerr-Newmann geometry. While a full scalarized black hole induced by parity violation is still lacking, it is crucial to understand the full scalarization dynamics by constructing the full scalarized black hole with parity-violating interactions, which remains a challenging problem \cite{nl2}. The main issue in this problem is that one has to solve the dynamical equation that may contain higher time derivatives in some specific parity-violating theories. For example, in the Chern-Simons modified gravity \cite{Alexander:2009tp}, the dynamical equation contains the third-order time derivative which results in the presence of ghost modes that generically render the theory pathological. In this sense, the parity-violating theories that contain high-order derivative terms can only be treated as effective field theories. This is also the reason why we only consider the dynamics in the ``decoupling limit." Recently, several parity-violating theories which are expected to be healthy have been proposed---for example, the Nieh-Yan modified teleparallel gravity \cite{Li:2020xjt, Li:2021wij} and symmetric teleparallel gravity with parity violations \cite{Li:2021mdp}. These particular parity-violating theories do not contain high-order derivative terms and do not have any ghost degree of freedom. In addition, a ghost-free parity-violating scalar-tensor theory, as an extension of the Chern-Simons modified gravity, has also been proposed \cite{Crisostomi:2017ugk}. It is interesting to explore the full dynamics of the scalar field and seek any possible full scalarized black hole solutions in these parity-violating theories. We expect to come back to this issue in our future works.

\section*{ACKNOWLEDGEMENTS}

This work is supported by the National Key Research and Development Program of China under Grant No.~2020YFC2201503, the Zhejiang Provincial Natural Science Foundation of China under Grants No.~LR21A050001 and No.~LY20A050002, and the National Natural Science Foundation of China under Grants No.~12275238, No.~11975203, and No. 12075207, and the US NSF Grant, No.~PHY2308845.

\end{document}